\documentstyle[12pt]{article}

\begin{document}
\begin{titlepage}

\title{Actions for signature change
         \thanks{
         Work supported by the Austrian Academy of Sciences
         in the framework of the ''Austrian Programme for
         Advanced Research and Technology''.}}

\author{Franz Embacher\\
        Institut f\"ur Theoretische Physik\\
        Universit\"at Wien\\
        Boltzmanngasse 5\\
        A-1090 Wien\\
        \\
        E-mail: fe@pap.univie.ac.at\\
        \\
        UWThPh-1995-1\\
        gr-qc/9501004
        }
\date{}

\maketitle

\begin{abstract}
This is a contribution on the controversy about junction conditions
for classical signature change. The central issue in this debate is
whether the extrinsic curvature on slices near the hypersurface of
signature change has to be continuous ({\it weak} signature change)
or to vanish ({\it strong} signature change).
Led by a Lagrangian point of view, we write down eight candidate
action functionals $S_1$,\dots $S_8$ as possible generalizations
of general relativity and investigate to what extent each of
these defines a sensible variational problem, and which junction
condition is implied. Four of the actions involve an integration
over the total manifold. A particular subtlety arises from the
precise definition of the Einstein-Hilbert Lagrangian density
$|g|^{1/2} R[g]$.
The other four actions are constructed as sums of integrals over
singe-signature domains. The result is that {\it both} types of
junction conditions occur in different models, i.e. are based on
different first principles, none of which can be claimed to
represent the ''correct''
one, unless physical predictions are taken into account.
{}From a point of view of naturality dictated by the variational
formalism, {\it weak} signature change is slightly favoured over
{\it strong} one, because it requires less {\it \`a priori}
restrictions for the class of off-shell metrics. In addition, a
proposal for the use of the Lagrangian framework in cosmology is
made.
\medskip

PACS-numbers: 04.20.Cv, 04.50.+h, 04.90.+e
\end{abstract}

\end{titlepage}

\section{Introduction}

The Euclidean path integral approach to quantum cosmology
\cite{HartleHawking}--\cite{Hawking2}
has led to the study of metrics that change their
signature type (from Euclidean to Lorentzian) at a
hypersurface $\Sigma$ (which is spacelike with respect
to the Lorentzian side). Originally, metrics emerging in a
WKB-approximation of quantum
gravity as ''real tunneling geometries'' were studied
\cite{GibbonsHartle}--\cite{Hayward2}.
The junction condition in this case is that the extrinsic
curvature at $\Sigma$ vanishes, when computed within the
embedding from either side (or appropriate generalizations
thereof when matter fields are added, typically
$\partial_t \phi=0$ at $\Sigma$ with $\partial_t$ the affine
comovig parameter
derivative).
We will refer to this type
of junction condition as {\it strong} signature change.
\medskip

Later, is was proposed that a classical change of signature type
is possible within a more relaxed framework as well,
with junction condition requiring the extrinsic curvature on slices
near $\Sigma$ to be continuous
\cite{EllisSumeruketal}--\cite{Ellis},
hence possibly non-zero, as opposed to the older approach
(the corresponding generalization to additional matter fields
being typically $\partial_t \phi =$ continuous). We denote
this situation as {\it weak} signature change. For some
work which has been done within this approach,
see e.g. Refs. \cite{DereliTucker}--\cite{FE3}.
\medskip

Since then, many aspects of these two types of junction
conditions have been studied (see for example, in addition
to the works quoted above, Refs.
\cite{KossowskiKriele1}--\cite{CarforaEllis}),
and occasionally the question which of these is the ''correct'' one
is adressed (more or less explicitly).
Regrettably, the different points of view are sometimes not
very clearly based on different underlying assumptions.
This has led to mutual accusation of ''errors'' among researchers
to an extent that seems to rather supply misunderstandings than to
clarify matters. The present piece of work is devoted to
the study of different assumptions (leading to different junction
conditions) from a Lagrangian point of view, hence to the
clarification of the question {\it which} first principles lead to
{\it which} type of junction condition.
\medskip

One plausible point of view (taken e.g. by Hayward
\cite{Hayward1},\cite{Hayward3})
is to consider Einstein's field equations
($R_{\mu\nu}=0$ in a distributional sense) as the fundamentals
(first principles) of a
model of classical signature change. (We will only consider
the {\it vaccum} case here -- except for Section 9,
where we add a cosmological constant --, as the inclusion of
additional
matter fields will not at all affect the principal arguments and
results given). To be more precise: by $R_{\mu\nu}$ we mean
the standard Ricci tensor as computed using the standard
formulae as e.g. given in Ref. \cite{MTW} (whose conventions
we are using) and inserting a metric that displays a change of
signature type on a hypersurface $\Sigma$.
It has been shown in Refs.
\cite{Hayward1},\cite{KossowskiKriele1}
that the resulting Einstein equations {\it may be interpreted}
in terms of generalized functions (distributions) and yield the
junctions conditions of
{\it strong} signature change as their $\delta$-function part
(see also equations (\ref{2.7})--(\ref{2.9}) below).
We will call these the ''full'' Einstein equations in order
to avoid confusion with the regular, single-signature
(Lorentzian or Euclidean) Einstein equations that have
to be satisfied off $\Sigma$ anyway.
\medskip

However, there is another, perfectly legitimate point
of view, namely the Lagrangian one. May classical signature change
be described within a proper variational framework, i.e.
is there a generally covariant action functional $S$ such that
$\delta S=0$ has signature type changing metrics as solutions?
In some sense, this point of view is even more fundamental
that the field equations approach, because it gives
quantization as well as coupling to realistic field
therories a conceptually firm basis. The former is of
course only important if quantisation is a goal. This
seems to be one source of misunderstandings: Shall we
quantize ''classical signature change'', as was envisaged
in Ref. \cite{Martin}, and expect oscillating WKB-solutions
even in the Euclidean domain, or shall we consider
''classical signature change'' as a classical model of
quantum tunneling in quantum cosmology
(that needs not be quantized at all)? Some authors
(see e.g. Ref.\cite{HellabyDray}) explicitly emphasize that their
intention is {\it not} a classical version of quantum
cosmology, and these statements seem to have been
overlooked by others. Here, we are just interested in
possible generalizations of general relativity that admit
a change of signature, hence we do {\it not} favour
one of the two types of junction conditions, and
we do {\it not} pre-assume any relation of such a model to quantum
cosmology (except for some speculations in Section 9).
\medskip

Let us note as a starting point that we have several
assumptions for first principles at our disposal
(based on Einstein's equations and based on
action principles), and that
we do not know (so far) which of them is the ''correct'' one.
We are just interested in the consequences of each choice.
\medskip

Although sometimes action functionals are written down
in the recent literature on signature change
(see e.g. Refs. \cite{Hayward1},\cite{KernerMartin}),
their variations and hence their implications
have, to my knowledge, not been worked out properly so far.
Maybe, some particular features that arise here have
simply been overlooked. One such feature is, as we will
see in Section 2, the possibility that
\begin{equation}
\delta S_{EH} = \int_{\cal M}d^n x\,|g|^{1/2}\, G_{\mu\nu}[g]
          \delta g^{\mu\nu} + {\rm additional\,\,terms}\, ,
\label{1.1}
\end{equation}
where $S_{EH}$ is the standard Einstein-Hilbert
action with a signature changing metric inserted
(although this provides some subtleties to which
we will return later on),
and the additional terms contain generalized functions
with support on $\Sigma$. Thereby, the variations vanish
outside some domain ${\cal N}$, but not necessarily
on $\Sigma$. (The variational prescription actually requires
$\cal N$ to be essentially arbitrary. The appearance of the
additional terms in (\ref{1.1})
may of course be circumvented by choosing
$\cal N$ to have empty intersection with $\Sigma$,
but this would not lead to a Lagrangian derivation of
signature change).
\medskip

{}From standard (Lorentzian signature) general relativity one is
used to the field equations $G_{\mu\nu}=0$ (or $R_{\mu\nu}=0$)
being equivalent
to the (variational) principle that the
Einstein-Hilbert action is stationary, $\delta S_{EH}=0$.
However, if the metric is allowed to change its signature type,
{\it this is not necessarily the case}, as (\ref{1.1}) shows.
As we will see in Section 2, the reason
for this fact is
a possible ''mismatch'' between the integrand (i.e. the
Lagrangian density) and the volume element.
However, the particular form of the
''additional terms'' in (\ref{1.1}) depends on the class
of admissible metric variations as well as on the
precise way the product between the Ricci scalar and the
volume element is defined, and there is one way of doing
it such that they vanish.
In this way, we will arrive at two possible
definitions of the Einstein-Hilbert action for
signature changing metrics.
Moreover, in standard
general relativity, one may subtract a boundary term from
the Einstein-Hilbert action
(it is sometimes called the Gibbons-Hawking boundary term;
see Ref. \cite{GibbonsHawking} for the appearance of this
modified action in quantum gravity). In the case of
signature change, such a modification ensures all
integrations to be performed over regular functions only,
but changes the class of admissible metric variations
and seems to have greater impact on the character of the model
than in standard general relativity.
\medskip

This amounts to re-examine the perspectives for the use
of variational principles when a change of signature is
envisaged. Let us provide some arguments
why $S_{EH}$ (which is, as mentioned,
not even a unique concept) is {\it not the only possible} action
functional.
\medskip

The main thing we can be sure about is that for the case
of a purely Lorentzian metric, the action functional shall coincide
with one of the standard actions for general relativity
($S_{EH}$ or $S_{EH}$ with the Gibbons-Hawking boundary term
removed). Furthermore, the action functional for the purely
Euclidean case shall
be of the same type (the Euclidean Einstein-Hilbert action).
However, here we have the choice of an additional {\it sign}
in front of it. There is no physical {\it \`a priori}
reason that would forbid us to assume the Euclidean
part of the action to appear with a minus sign
relative to the Lorentzian part. A formal justification
of such an assumption is that the metric signature is
a coordinate-independent (hence ''covariant'') concept and
thus might in principle provide a constituent of an action
integral.
\medskip

A further issue where the possibility of a {\it choice} appears
is the question how to treat the integration across the
signature change hypersurface $\Sigma$. In view of
the distributional character of the full Ricci scalar at $\Sigma$,
one might split the action functional into two contributions,
each being defined as an integral over the respective
single-signature domain.
The advantage of this approach is that the integrals
appearing are perfectly well-defined, and no distibutions
have to be dealt with.
There is nothing exotic with such a concept. On the contrary, it
applies to standard (Lorentzian signature) general relativity
as well: Assume that any hypersurface cuts the total space-time
manifold into two pieces, and postulate a variational principle
defined by the sum of the respective Einstein-Hilbert actions
of the two
pieces. Then, under mild continuity and differentiability
conditions for the total class of (off-shell) metrics, one
recovers Einstein's field equations.
Let us call this the ''additivity property''.
\medskip

Here a comment on our use of the words ''on-shell'' and
''off-shell'' (that originally referred to the mass shell
condition for particles in special relativity) is in place:
By ''off-shell'',
we denote those metrics which are considered from the outset,
before field equations or variational
principles are imposed. The variations are performed
{\it within} this class of metrics. The physical solutions
(i.e. the solutions of field equations or variational
principles) are called ''on-shell''. The precise definition
of the class of off-shell metrics depends on the model
considered (this definition is actually a part of the model,
along with the formal action integral) and is usually
made such that the action $S[g]$ is well-defined and
the variational principle $\delta S[g]=0$ admits nontrivial
solutions. However, there is always a certain
freedom to modify models by redefining the class of off-shell
metrics.
\medskip

Thus, combining these possibilities for action functionals,
we arrive at eight candidates. If $\epsilon$ denotes
the metric signature ($\epsilon=1$ for a Lorentzian metric and
$\epsilon=-1$ for a Euclidean one), these eight candidates are
constructed as follows:
$S_1\equiv S_{EH}$ is obtained by integration
over the total Ricci scalar $R[g]$, $S_2$ by integration
over $\epsilon[g] R[g]$.
Thereby, the absolute value of the metric determinant
$|g|^{1/2}$ is defined as a (continuous) object
and thereafter multiplied by $R[g]$ and
$\epsilon[g]R[g]$, respectively.
$S_3$ and $S_4$ are anologous constructions, but with the
product $|g|^{1/2}\times R[g]$ defined in a different way.
We denote all these four candidates as ''singular actions'',
because the integrands have to be interpreted in terms
of generalized functions (if possible).
$S_5$ is the integral over $R[g]$, but
broken into two regular contributions from the Lorentzian and the
Euclidean domain, and $S_6$ is the corresponding
expression using $\epsilon[g] R[g]$. Finally $S_7$ and
$S_8$ are the regular two-domain constructions based on
$R[g]$ and $\epsilon[g]R[g]$, but with appropriate
boundary terms subtracted. The last four candidates will be
called ''regular actions''.
\medskip

The division into singular and regular actions will prove important
as the former are plagued by ill-defined nonlinear combinations
of distributions. Our attempts to overcome these problems are
somewhat
heuristic and unsatisfactory, and in all cases they are successfull,
we will end up effectively within the framework of regular actions.
This has of course some impact on our notion of ''naturality'', and
among our results maybe the most appealing one will be a relation
between $S_8$ and the ''additivity property''.
\medskip

So far, all eight candidate actions have been introduced
{\it formally} (except for the subtleties
distinguishing $S_1$ and $S_2$ from $S_3$ and $S_4$).
It is clear that {\it off} the hypersurface
$\Sigma$ of signature change, they just procuce the
single-signature (Lorentzian or Euclidean) Einstein equations.
The crucial point is what they predict at (more
precisely: across) $\Sigma$.
In order to examine their ability to
describe signature change within a reasonable variational
framework, we have to adress several questions to each of them:
(i) In which sense is it well-defined?
(ii) What is the class of off-shell metrics, i.e. the set of metrics
{\it within which} the variations are performed? Since all
variations are infinitesimal, this question may
in a sloppy way be rephrased as:
What is the class of off-shell variations?
(By ''off-shell'', we mean ''not necessarily a solution'',
whereas ''on-shell'' denotes the solutions of the variational
principle).
(iii) Which type of junction condition follows?
These questions will be answered for all eight candidate
actions.
\medskip

Our results may be summarized as follows:
The answers to the questions posed
above will essentially be that the actions
based on $R[g]$
($S_1$, $S_3$, $S_5$ and $S_7$) are associated with {\it strong},
and those based on $\epsilon[g] R[g]$
($S_2$, $S_4$ $S_6$ and $S_8$) are associated with {\it weak}
signature change, although on a very different level
of well-posedness and naturalness within the usual spirit
of Lagrangian formulations. The main problems arise from the
fact that for some cases $\delta S=0$ admits nontrivial
solutions only if severe restrictions on the class of off-shell
metrics are imposed. Disappointingly, $S_1\equiv S_{EH}$
(in the interpretation that $|g|^{1/2}$ is a well-defined
object by its own before multiplied by $R[g]$)
defines a proper variational problem only if the
{\it strong} junction conditions are imposed from the outset
on the class of off-shell metrics (which is clearly
an unpleasant feature within a variational formulation).
A similar problem is encountered with $S_2$, $S_4$ and $S_5$
(although
in the two former cases an {\it ad hoc} but not very beautiful
regularization scheme might save
the situation). Altogether, these results seem to indicate that
the singular actions $S_1$, $S_2$, $S_4$ and $S_5$ are ruled out.
The singular action $S_3$
(which turns out to be essentially the same as $S_7$) defines
a model for
{\it strong} signature change, once a differentiability condition
is imposed on the class of off-shell metrics.
(Such a condition is acceptable although not very appealing).
This model then leads directly to the
full (distributional) Einstein equations. Its peculiar feature
is that the Lagrangian density is not constructed as
an ordinary product $|g|^{1/2}\times R[g]$.
(To what extent this is a ''disadvantage'' is a matter
of philosophy).
The candidates $S_6$ and $S_7$ ($\equiv S_3$) define sensible
models only if
the off-shell metrics satisfy a differentiability
condition, and hence give rise to more or less
acceptable models for {\it weak} and {\it strong}
signature change, respectively. The winner is $S_8$, which
only needs a continuity condition for the off-shell metrics
and leads to {\it weak} signature change. This last
action is a natural generalization of what one uses in
Euclidean path integral quantum cosmology
\cite{HartleHawking},\cite{Hawking2},
i.e. with the usual Gibbons-Hawking boundary term
\cite{GibbonsHawking} subtracted,
and with an additional minus sign for the Euclidean sector.
All these results are valid in an arbitrary
number of space-''time'' dimensions $n\geq 3$.
\medskip

In terms of a (very simplifying) toy model, the actions
based on $R[g]$ correspond to the Lagrangian
(let $\epsilon\equiv$ ${\rm sgn}(x)$)
\begin{equation}
{\cal L}_1^{\rm toy}= \frac{1}{2}\, \epsilon\, \dot{x}^2 - U(x),
\label{1.2}
\end{equation}
whereas the actions based on $\epsilon[g]R[g]$ correspond
to the Lagrangian
\begin{equation}
{\cal L}_2^{\rm toy}\equiv \epsilon\,{\cal L}_1^{\rm toy}
= \frac{1}{2}\, \dot{x}^2 -\epsilon\, U(x).
\label{1.3}
\end{equation}
The junction conditions are (if the sign changes at
$t=0$) $\dot{x}=0$ ({\it strong}) and
$\dot{x}(-0)=\dot{x}(+0)$ ({\it weak}), respectively.
\medskip

As a by-product of the pattern of signs emerging, we will
see that the regular candidate actions $S_6$ and $S_8$
(both associated with {\it weak} signature change)
may be formulated for the single-signature case as well
and precisely provide the statement that the single-signature
Einstein-Hilbert action may be broken into a sum of integrals
and still produces Einstein's field equations (as was
mentioned above as the ''additivity property'').
Thus, from the point of view of the sign
structure, the invariant $\epsilon[g]R[g]$ seems to provide
a variational principle in a way as natural as $R[g]$.
\medskip

This may be illustrated in terms of the toy model given above.
The natural
action to consider is (let $\epsilon_{\pm}(t)$ be
${\rm sgn}(x(t))$ for $t>0$ and $t<0$, respectively)
\begin{equation}
S^{\rm toy}=\epsilon_{-} \int_{t_{\rm min}}^0 dt\,
            {\cal L}_1^{\rm toy}
         +  \epsilon_{+} \int_0^{t_{\rm max}} dt\,
            {\cal L}_1^{\rm toy}.
\label{1.4}
\end{equation}
Then the sign combinations with $\epsilon_{+}=-\epsilon_{-}$
produce the analogue of {\it weak} signature change, whereas
the sign combinations with $\epsilon_{+}=\epsilon_{-}$
just illustrate the ''additivity property'' ({\it weak}
signature change thus appearing as a generalization
thereof). We should
add that this toy model displays only part of the structure
encountered in gravity (namely as far as the lapse and shift
degrees of freedom, a divergence term and some further
subtleties are ignored).
\medskip

In this way one might, contrary to the most approaches
to signature change, get the
intuitive feeling that the Ricci scalar $R[g]$ as it is
used in the Einstein-Hilbert action for standard general relativity
is in fact $\epsilon[g]R[g]$, but with $\epsilon[g]=1$ inserted,
since the metric is Lorentzian ''today''. Hence, there is no reason
to consider those actions which display a non-standard sign in the
Euclidean domain as less natural.
This concludes the summary of the results obtained.
\medskip

There may be, however, yet other possibilities for
action functionals that we have
not considered here (as e.g. the inclusion of further
boundary integrals that would act like a gauge-fixing term,
enforcing the {\it strong} junction conditions). Moreover,
we have not included matter fields (except for a cosmological
constant in Section 9) because their presence would
neither change our arguments nor the structure of the results.
In particular the generalization to bosonic matter
fields seems to be staightforward.
\medskip

Let us add the conceptual remark that in a proper Lagrangian
framework the hypersurface $\Sigma$ has the status of a
{\it variable} too, and the metric variations $\delta g_{\mu\nu}$
have to be supplemented by variations of the location
of the hypersurface, symbolically denoted as $\delta\Sigma$.
However, it will turn out in Section 8 that
for the most reasonable candidates, $S_7$ and $S_8$,
this aspect does not produce
any new information, so that in the main part of this paper
$\Sigma$ is held fixed (and described by the equation
$x^0\equiv t=0$).
As a consequence of $\Sigma$ being part of the dynamical
variables, the Lagrangian formulation does not tell us
whether a classical change of signature actually {\it will}
occur. Consider e.g. a Lorentzian metric that develops all
necessary conditions for a signature change at some spacelike
hypersurface $\Sigma$. Then {\it two} solutions are possible:
one with and one without signature change. This is a true
non-uniqueness of solutions to the Lagangian formulation,
and it is already present in an approach based on field
equations and junction conditions. Presumably only the
quantization or an {\it ad hoc} assumption can remove it.
In Section 9, we will give an example for such an
assumption (it essentially states -- in a cosmological framework
-- that signature change will always occur whenever possible,
and thus render the metric Euclidean/Lorentzian in the sectors
of superspace that are usually referred to as
Euclidean/Lorentzian).
This type of non-uniqueness is not an obstruction, and
a comparable thing happens in classical string
theory \cite{GreenSchwarzWitten}, where the
world-sheet topology between a given initial and final
configuration is not unique, due to ''unpredictable''
splitting and joining effects of the pieces of the string.
\medskip

In order to avoid possible confusion with another
''classification'' of methods, we should make a further,
more technical remark.
There have been developed two notions of classical signature change,
labelled as the ''continuous'' and the ''discontinuous'' approach
(see e.g. Refs. \cite{EllisSumeruketal},\cite{KossowskiKriele1}).
However, the difference between these approaches is to some
extent only a coordinate transformation
\cite{KossowskiKriele1},\cite{KrieleMartin}. The prototype
situation is given by the two forms
\begin{equation}
-{\rm sgn}(t)\, dt^2 = -\,\tau\, d\tau^2
\label{ccc}
\end{equation}
for the pure ''time''-part of the metric, with $t$ being the
affine comoving parameter and $\tau$ the corresponding
''continuous approach'' coordinate. Hence,
${\rm sgn}(t)={\rm sgn}(\tau)$ and $|t|=(2/3)|\tau|^{3/2}$.
The distinction between them will not be important at all.
This comes from the fact that only the limits
of quantities (like the metric) as computed within the
embedding from either side are relevant for the Lagrangian
point of view (together with the distributions
that are generated by differentiating discontinuous functions
with respect to $t$),
but we never need to specify, say,
the particular value of $g_{00}$ at
$\Sigma$ (i.e. at $t=0$), even if $g_{00}$ is discontinuous. As a
consequence, a simple coordinate transformation will translate
any statement made in the ''discontinuous'' approach into one made
in the ''continuous'' approach, and {\it vice versa}.
We prefer to use the language of the ''discontinuous''
approach, because Gaussian coordinates ($g_{00}=\pm 1$, $g_{0i}=0$)
are particularly easy to handle.
Note however that the two coordinates $t$ and $\tau$ are
related to two different notions of ''differentiability''.
The existence of $\partial_t f$ is not equivalent to the
existence of $\partial_\tau f$ at $t=\tau=0$, for any function $f$.
In the Lagrangian framework we will not presuppose any such
condition for the physical variables (in particular the
spatial metric), and the natural notion of differentiability
associated with this approach will be the one with respect
to $t$. From the Lagrangian point of view there is nothing special
about $\tau$, whereas $t$ has a preferrred status as the local
orthogonal metric distance from $\Sigma$.
\medskip

The paper is organized as follows: In Section 2, we give
some preliminaries, in particular the variations of some
Lagrangian density expressions, when signature change
is allowed. In Section 3, the definition of the
list of candidate actions is given.
Section 4 is devoted to the study of the singular actions $S_1$
and $S_2$, Section 5 is concerned with the singular actions
$S_3$ and $S_4$. The regular actions $S_5$ and $S_6$
(using the total single-signature
Ricci scalars) are considered in Section 6, the regular actions
$S_7$ and $S_8$ (being based on the subtraction of a surface
integral) are dealt with in Section 7. In Section 8, we include
the variation of the
hypersurface $\Sigma$. Section 9 deals with a coordinate
dependent but instructive form of $S_7$ and $S_8$ and
a proposal for a resolution of the non-uniqueness problem
within the frawework of cosmology. Finally,
Section 10 contains some concluding remarks.
\medskip

\section{Variations}
\setcounter{equation}{0}

In order to define a convenient $(n-1)+1$ split formalism with
coordinates $x^\mu\equiv(t,x^i)\equiv(t,{\bf x})$, assume the
choice of a slicing $\Sigma_t$ of hypersurfaces defined by constant
values of $t$, such that a signature change occurs at
$\Sigma\equiv\Sigma_0$. (One may in fact consider several
signature changes, in which case it is reasonable to parametrize
their hypersurfaces as $\Sigma_j\equiv\Sigma_{t_j}$).
Assume the hypersurface(s) of signature change to be spacelike
with respect to the Lorentzian side(s). Furthermore, let $n^\mu$
be the (unique) vector field orthogonal to the slices,
''future'' directed with respect to $t$ ($n^0\geq 0$) and having
the square $n^\mu n_\mu = - \epsilon$.
Later on, we shall choose
$\epsilon=1$ for the Lorentzian and $\epsilon=-1$ for the Euclidean
domains.
However, in view of the distributional character
of some quantities appearing, we
treat $\epsilon$ as an arbitrary function of the ''time''
coordinate, $\epsilon\equiv\epsilon(t)$, in all general formulae
derived here. Then $n_\mu=$ $-\epsilon N\partial_\mu t$
(or $n^0\equiv$ $ n^\mu \partial_\mu t =$ $ 1/N$) defines
the lapse function $N\geq 0$. The induced metric (first
fundamental form) on $\Sigma_t$
is given by $h_{\mu\nu}=$ $g_{\mu\nu}+$ $n_\mu n_\nu/\epsilon$,
the extrinsic curvature (second fundamental form) by
$K_{\mu\nu} =$ $ - h_\mu^\rho h_\nu^\sigma \nabla_\rho n_\sigma$.
This is actually a sloppy notation, and in order to be more
precise we mention that the ''true'' extrinsic curvature
is $K_{\mu\nu}/|\epsilon|^{1/2}$. It is a ''covariant'' object,
and it proves useful if one tries to reproduce the results of this
paper in a manifestly ''continuous'' approach (which amounts to set
$\epsilon(t)=t$). We will rather use a manifestly ''discontinuous''
language in which $\epsilon$ (after a sequence of steps during
which it is treated as arbitrary function) is set equal to
a step function (preferably $\pm 1$), and $N$ is continuous.

The spatial components of $n^\mu=(1/N,-N^i/N)$ (or the
components $h_{0i}=N_i$) define the shift vector. The metric
reads
\begin{eqnarray}
ds^2 &=& - \epsilon N^2 dt^2
       + h_{ij} (dx^i + N^i dt)(dx^j + N^j dt)
\label{2.1b}\nonumber\\
     &=& (- \epsilon N^2 + N^i N_i) dt^2 + 2 N_i dt dx^i
                +h_{ij}dx^i dx^j,
\label{2.1}\\
\nonumber
\end{eqnarray}
where the induced metric $h_{ij}$ appears as an $(n-1)$-object and
serves (togeher with its inverse $h^{ij}$) to raise and
lower spatial indices (e.g. $N_i = h_{ij} N^j$). The components
$N$, $N_i$ and $h_{ij}$ are functions of all coordinates.
Note that discontinuous $\epsilon$
(although producing $\delta$-distributions
wherever $\dot\epsilon$ appears) does not necessarily imply
discontinuous metric components (as the example
$\epsilon(t)={\rm sgn}(t)$, $N(t)=|t|^{1/2}$ shows: this is
equivalent to $\epsilon(t)=t$, $N(t)=1$ and thus another
way to implement the ''continuous'' approach).
The spatial version of the extrinsic curvature is
\begin{equation}
K_{ij} = \frac{1}{2N}\bigg(
      N_{i|j} + N_{j|i} - \partial_t\, h_{ij} \bigg),
\label{2.2}
\end{equation}
where ${}_{|}$ denotes the covariant derivative with respect to
$h_{ij}$. Furthermore, we define $K=K^i_i\equiv$ $h^{ij}K_{ij}$.
Note that $\epsilon$ has not been constrained so far (apart
from $\partial_i \epsilon=0$). Only if $\epsilon=\pm 1$,
the above construction coincides with the standard
$(n-1)+1$ split formulation
(see e.g. Refs. \cite{MTW},\cite{DeWitt} for the
Lorentzian signature case).
\medskip

The determinant
of (\ref{2.1}) is $g\equiv {\rm det}(g_{\mu\nu})=$
$ - \epsilon N^2 h$, with $h\equiv {\rm det}(h_{ij})$.
The volume element for the integration in the action
shall be defined as $|g|^{1/2}\,d^n x$. Putting
$\epsilon=\pm 1$ in the respective domains amounts to
set $|\epsilon|=1$ and hence $|g|^{1/2}=N h^{1/2}$.
This definition aims at the construction of the
volume element as a well-defined quantity by its own.
(This construction of a volume element is quite natural and
is occasionally written down explicitly for signature
changing metrics, see e.g. Ref. \cite{DrayHellaby}).
Its prefactor is continuous (as long as $N$ and $h^{1/2}$ are),
and an integration is defined by multiplying this
object by the integrand (which may be a distribution and
is considered as a well-defined object by its own as well).
The Einstein-Hilbert action is such an integral,
the integrand being the Ricci scalar $R[g]$ (a
combination of ordinary functions and $\delta$-distribution;
see equation (\ref{2.3}) below). This leads to the version
$S_1$ of the Einstein-Hilbert action.
\medskip

However, there is an other possibility that will lead
to a different version of the Einstein-Hilbert action
(namely to $S_3$).
It consists of considering the ''product''
$|g|^{1/2}\, R[g]$ as the result of a limiting process,
in which first $\epsilon$ is treated formally as if it were
an arbitrary function, and in a second step the resulting formal
expression for $|g|^{1/2}\, R[g]$ is given some sense.
This in turn amounts to define the volume element
formally as $|\epsilon|^{1/2}\,N h^{1/2}\, d^n x$.
When varying the Einstein-Hilbert action defined in this
way, derivatives $\partial_t |\epsilon|$ appear.
Although this is zero for a step function in the usual
sense of distributions, a complication arising
is that it has to be multiplied by further $\epsilon$-terms,
so that the final expression is not well-defined
as it stands and has to be {\rm interpreted}.
One interpretation is to insert $|\epsilon|=1$, which
renders $\partial_t |\epsilon|=0$ (hence reproduces
the choice $|g|^{1/2}=N h^{1/2}$), but another
interpretation is to identify
$(\partial_t |\epsilon|)/|\epsilon|$ with
$(\partial_t \epsilon)/\epsilon$, and this will
provide a different action.
Hence, we keep in mind that we have two
(different) definitions for volume integrals at hand.
In both schemes it is important not to use identities like
$1/\epsilon=\epsilon$ or $\epsilon^2=$ $|\epsilon|=1$
too early. As an example, $\partial_t(1/\epsilon)$ is to be
evaluated as $-\dot{\epsilon}/\epsilon^2$.
The reason for these subtle issues is that distributions occur
in a non-linear manner.
We decide to keep such
expressions as they stand (in particular distinguishing between
$\epsilon$ and $1/\epsilon$) in all formulae throughout the whole
paper, unless obvious or stated otherwise. Occasionally, one
has the freedom to interpret ill-defined objects
in different ways: the volume element problem is an example
of such a case. (Rephrased in the quantum field theory jargon
this means that one may apply different ''regularization''
schemes).
All these subtleties (e.g. in defining the behaviour of the
Lagrangian density across $\Sigma$) are of course irrelevant
for the regular actions $S_5$,\dots $S_8$, who
avoid any problems of this sort from the outset.
\medskip

Locally, one can always choose coordinates such that $N=1$ and
$N_i=0$ (thus $x^0\equiv t$ being the
local orthogonal metric distance from $\Sigma$, or --
in other words -- the ''affine comoving parameter'').
Such coordinates are referred to as Gaussian ones, and
(once $\Sigma\equiv\Sigma_0$ is given)
they are unique up to $t$-independent
spatial transformations $x^i_{\rm new} = X^i({\bf x})$. In a
Gaussian coordinate system, we require $h_{ij}$ to be
strictly positive, hence $h>0$.
\medskip

The Ricci scalar with respect to (\ref{2.1}) is given by
(cf. Refs. \cite{MTW},\cite{DeWitt})
\begin{eqnarray}
R[g]&=&R[h]+\frac{1}{\epsilon}
   \bigg( K^{ij}K_{ij}+K^2\bigg)
  -\frac{2}{N} N^{|i}{}_{|i}
\label{2.3a}\nonumber\\
  &{}& -\,\frac{\dot\epsilon}{\epsilon^2 N}\, K
  + \frac{1}{N}\bigg(\partial_t-N^i \partial_i\bigg)
      \,\bigg( -\frac{2}{\epsilon}\, K\bigg),
\label{2.3}\\
\nonumber
\end{eqnarray}
where again $\epsilon\equiv \epsilon(t)$ has been treated as an
arbitrary function. If $\epsilon=\pm 1$ is a step function,
one could insert $\epsilon^2=1$ and $\dot{\epsilon}$
a $\delta$-function.
(Let us repeat that some caution is necessary here: one is not
allowed to insert the same step function for $1/\epsilon$ in
the last expression before the ''time''-derivative is carried out).
In a coordinate system with zero shift and $N\equiv N(t)$,
our expression for $R[g]$ coincides with the one given by
Kossowski and Kriele (in the appendix of
Ref. \cite{KossowskiKriele1}).
Also, these authors adopt -- in part of their paper --
the ''discontinuous'' language and proceed along similar lines
when computing the energy-momentum tensor. Clearly, $R[g]$
is only a well-defined distribution if $K$ (which appears
multiplied by $\dot{\epsilon}/\epsilon^2$) is continuous.
This implies a differentiability condition for the class of
off-shell metrics for all models using the full Ricci scalar and
an integration over the total manifold. We will impose such
restrictions in detail later on, when discussing the particular
models. However, we note that just this type of ill-definedness
of quantities for a large class of metrics provides the major
drawback for all singular actions.
\medskip

In order to proceed, we multiply $R[g]$ by $N h^{1/2}$ (which
is $|g|^{1/2}$ in the interpretation that $|\epsilon|=1$
is to be inserted). Thus, the integrand for the action $S_1$
(and a basic quantity from which the other actions
are constructed as well)
\begin{equation}
{\cal L} = N h^{1/2}\, R[g]
\label{2.4}
\end{equation}
is given by
\begin{eqnarray}
{\cal L}&=&N h^{1/2}\, R[h]+\frac{1}{\epsilon}\, N h^{1/2}\,
   \bigg( K^{ij}K_{ij}-K^2\bigg)
  -\frac{\dot\epsilon}{\epsilon^2} \,h^{1/2}\, K
\label{2.5a}\nonumber\\
   &{}& +\, \partial_t \,\bigg( -\frac{2}{\epsilon}\,
       h^{1/2}\, K\bigg)
   + \partial_i \bigg(\frac{2}{\epsilon} \, h^{1/2} \,N^i K
         - 2\, h^{1/2}\, N^{|i} \bigg).
\label{2.5}\\
\nonumber
\end{eqnarray}
In all integrations over the spatial coordinates, divergence
terms like the last contribution in (\ref{2.5}) will
be omitted.
The role of momenta in the canonical formulation is played
by the spatial tensor density
\begin{equation}
\widetilde{\pi}^{ij}=
   \frac{\partial}{\partial \dot{h}_{ij}}
     N\,h^{1/2}\,\bigg(K^{kl}K_{kl} - K^2\bigg)
   = h^{1/2} \bigg(K h^{ij} - K^{ij}\bigg)
\label{2.6}
\end{equation}
or its negative (according to the sign in front of $\cal L$
and the sign of $\epsilon$).
\medskip

In order to take a look at  Einstein's field equations
$R_{\mu\nu}=0$, we write down the components of the Ricci
tensor (for simplicity in a coordinate system in which
$N_i=0$)
\begin{eqnarray}
R_{00}[g]&=& N \partial_t K -\frac{\dot\epsilon}{2\epsilon} N K
       - N^2 K^{ij} K_{ij}
        + \epsilon N N^{|i}{}_{|i},\label{2.7}\\
R_{0i}[g]&=& N (\partial_i K - K^j_{i|j}),\label{2.8}\\
R_{ij}[g]&=&R_{ij}[h]+\frac{\dot\epsilon}{2\epsilon^2 N} K_{ij}
\label{2.9a}\nonumber\\
 &{}& -\,\frac{1}{\epsilon}\bigg(
          \frac{1}{N}\, \partial_t K_{ij} + 2K_i^l K_{lj}
                                            -K K_{ij} \bigg)
          - \frac{1}{N} N_{|ij}.
\label{2.9}\\
\nonumber
\end{eqnarray}
Before going on to the Lagrangian approach, we observe that
Einstein's equations decouple into a regular part
(the single-signature field equations) and a contribution
proportional to $\dot \epsilon$ (which is a $\delta$-distribution
at $\Sigma$, where $\epsilon$ is discontinuous). In Gaussian
coordinates, the latter reduces to
$\partial_t h_{ij}=0$ at $\Sigma$, hence the junction
conditions for {\it strong}
signature change. This fact provides the basis from which
the concept of {\it weak} signature change is usually
critizised as not satisfying ''the correct'' field
equations (see e.g. Ref. \cite{Hayward3}). Such criticism
is of course obsolet, once ''the'' field equations are
defined in terms of variational problems, both {\it strong}
and {\it weak} signature change being possible and
associated with different actions.
\medskip

The Lagrangian formulation amounts to compute variations of
the action due to infinitesimal changes {\it off} a given
metric. We will assume this given metric to be expressed
in terms of Gaussian coordinates. However, the variations
{\it off} this metric must be generic, i.e. they induce
nontrivial lapse and shift functions, although at an
infinitesimal level. Hence, we must use the full expression
(\ref{2.5}) when varying $\cal L$.
As independent variations we choose $\delta N$, $\delta N_i$
and $\delta h_{ij}$, and we use
\begin{equation}
A^{ij}= N^{i|j} - \,\frac{N^{|i}}{N}\,N^j
\label{2.10}
\end{equation}
as an abbreviation. For the sake of generality, we display
the full structure of $\delta {\cal L}$:
\medskip

{\bf Variation with respect to the lapse:}
\begin{equation}
\delta {\cal L} = {\cal E} \delta N
    +\frac{\dot \epsilon}{\epsilon^2 N} h^{1/2}\, K \delta N
   +\partial_t\,\bigg(\frac{2}{\epsilon N}\,
    h^{1/2}\, K \delta N\bigg)
   + \partial_i {\cal U}^i.
\label{2.11}
\end{equation}

{\bf Variation with respect to the shift:} The independent
variation is $\delta N_i\equiv$ $ \psi_i$, and we set
$\psi^i\equiv$ $ h^{ij}\psi_j$.
\begin{eqnarray}
\delta {\cal L} &=& {\cal E}^i \psi_i
        - \frac{\dot \epsilon}{\epsilon^2 N^2}\,
         h^{1/2}\, N^{|i} \psi_i
        +\partial_t\, \bigg(- \frac{2}{\epsilon N^2}\,h^{1/2}\,
                           N^{|i} \psi_i \bigg)
\label{2.12a}\nonumber\\
       &{}& +\,\partial_i {\cal V}^i.
\label{2.12}\\
\nonumber
\end{eqnarray}

{\bf Variation with respect to the induced metric:} As the
independend variation we take
$\delta h_{ij}\equiv$ $\psi_{ij}$. Then, indices are raised and
lowered by $h_{ij}$ and $h^{ij}$ (as e.g. $\psi^i_j\equiv$
$h^{ik}\psi_{kj}$).
\begin{eqnarray}
\delta {\cal L} &=& {\cal E}^{ij} \psi_{ij}
   +\frac{\dot\epsilon}{2\epsilon^2 N}\,h^{1/2}\,
      \bigg( \Big(N K + A^j_j\Big)\psi^i_i
           - 2 A^{ij} \psi_{ij} + h^{ij} \dot{\psi}_{ij} \bigg)
\label{2.13a}\nonumber\\
&{}&+\,\partial_t\,
        \bigg(   \frac{1}{\epsilon N}\,h^{1/2}\,
       \Big( A^j_j \psi^i_i + (N K^{ij} - 2 A^{ij})\psi_{ij}
        + h^{ij} \dot{\psi}_{ij}  \Big) \bigg)
\label{2.13b}\nonumber\\
  &{}&    +\, \partial_i {\cal W}^i.
\label{2.13}\\
\nonumber
\end{eqnarray}
In these formulae the ${\cal E}$'s denote the regular,
single-signature Einstein equations
(i.e. without $\dot\epsilon$-terms). The divergence terms (denoted
by $\partial_i {\cal U}^i$ etc.), contain expressions
involving $\dot\epsilon$, and
they belong to the sort of contributions we omit when
integrated over.
\medskip

Here, the formal reason for the appearance of $\dot\epsilon$-terms
{\it apart} from those contained in the
(full, i.e. distributional) Einstein equations
(cf. equation (\ref{1.1})) becomes clear. Normally, one expects
the metric determinant to absorb all such additional
contributions (this is the step from $\delta R[g]$, which we
have not displayed, to $\delta {\cal L}$). However, the
function $\epsilon(t)$ does not appear in $|g|^{1/2}=N h^{1/2}$
because its absolute value $|\epsilon|$ has been set $1$.
Clearly, as long as $\epsilon$ is treated as an arbitrary
function in $R[g]$, one expects the appearance
of anomalous $\epsilon$-terms in $\delta {\cal L}$,
hence a ''mismatch'' between the Lagrangian density and the
volume element. If $\epsilon$ is a step function
with values $\pm 1$, one might expect this ''mismatch''
to disappear. However, this is not the case! Hence, only
part of the $\dot\epsilon$-terms in $\delta {\cal L}$ are due to
the (full) Einstein equations as given by (\ref{2.7})--(\ref{2.9}).
This is the reason for the field equations point of view
being not necessarily equivalent to the Lagrangian point of view.
Many troubles with signature change seem to stem from this fact,
and it is unavoidable as long as the volume element in the action
is defined as an object by its own, namely $|g|^{1/2} \, d^n x$ with
$|\epsilon|=1$ inserted.
\medskip

Here we arrive at the second possibility for interpreting
the product $|g|^{1/2}\,R[g]$ that will lead us to the
version $S_3$ of the Einstein-Hilbert action. As already
mentioned above, we formally assume
$|g|^{1/2}=$ $|\epsilon|^{1/2}\,N h^{1/2}$ and define
the modified Lagrangian density
\begin{equation}
{\overline {\cal L}} = |\epsilon|^{1/2}\,{\cal L}.
\label{z.1}
\end{equation}
Treating $|\epsilon|^{1/2}$ as an arbitrary function, we obtain,
from (\ref{2.5}),
\begin{eqnarray}
{\overline {\cal L}}&=&|\epsilon|^{1/2} N h^{1/2}\, R[h]+
     \frac{|\epsilon|^{1/2}}{\epsilon}\, N h^{1/2}\,
   \bigg( K^{ij}K_{ij}-K^2\bigg)
  - {\cal F} \,h^{1/2}\, K
\label{z.2a}\nonumber\\
   &+&  \partial_t \,\bigg( -
     2\frac{|\epsilon|^{1/2}}{\epsilon}\, h^{1/2}\, K\bigg)
   + \partial_i \bigg(2 \,\frac{|\epsilon|^{1/2}}{\epsilon} \,
            h^{1/2} \,N^i K
         - 2 \,|\epsilon|^{1/2}\, h^{1/2}\, N^{|i} \bigg),
\label{z.2}\\
\nonumber
\end{eqnarray}
where
\begin{equation}
{\cal F} = \bigg(\, \frac{\partial_t \epsilon}{\epsilon}
                -  \frac{\partial_t |\epsilon|}{|\epsilon|}
                 \, \bigg)
      \,\frac{|\epsilon|^{1/2}}{\epsilon}.
\label{z.3}
\end{equation}
Hence, the net effect of the transition from ${\cal L}$ to
$\overline{{\cal L}}$ is (apart from the $|\epsilon|^{1/2}$-factors)
that $\dot{\epsilon}/\epsilon^2$ has been replaced by ${\cal F}$.
Inserting $|\epsilon|\equiv 1$ returns of course
$\dot{\epsilon}/\epsilon^2$. An alternative interpretation
is to ''regularize'' everything by looking at $\cal F$
as $\epsilon$ were an arbitrary (say, smooth) function. Then
${\cal F}(t)=0$ for all $t$ such that $\epsilon(t)\neq 0$. Whether
the zeros of a smooth version of $\epsilon(t)$ produces
a non-trivial distribution is a matter of interpretation, because
${\cal F}(t)$, as it stands, is not well-defined. Its two
contributions
produce bad singularities when $\epsilon(t)=0$. Thus a possible
interpretational scheme for an alternative version of the
Einstein-Hilbert action is to set ${\cal F}\equiv 0$ by hand.
This is what we will do here, in order to define
${\overline{\cal L}}$ unambigously.
However, note that this interpretations does not allow
$|g|^{1/2}$ and $R[g]$ to be objects by their own that are
just multiplied. The reason for this alternative Lagrangian
density to exist is the freedom in choosing the order in which
$\epsilon$ is {\it (a)} ''regularized'' and {\it (b)} set
equal $\pm 1$ in the various quantities. Another way of looking
at this is to insist on the Lagrangian density
${\overline{\cal L}}$ being the
ordinary product of $|g|^{1/2}$ (as used in the definition of
$S_1$) with ''the'' Ricci scalar. This amounts to propose a
modified definition for the Ricci scalar of a signature changing
metric by
${\overline{\cal L}}=$ $N h^{1/2}\,{\overline R}[g]$. It would
be interesting to study the properties of ${\overline R}[g]$
(in comparison to $R[g]$) in more detail than is possible here.
\medskip

The variations of ${\overline{\cal L}}$
proceed along the same lines as those given
above. In order to be brief, we just write down the
overall structure (obtained without setting ${\cal F}\equiv 0$,
for the moment)
\begin{equation}
\delta {\overline{\cal L}}={\overline{\cal E}}+{\cal F}{\cal Z}
  + {\cal G} +\partial_\mu {\cal X}^\mu,
\label{z.4}
\end{equation}
where ${\overline{\cal E}}$ stands for the single-signature
Einstein equations (without $\dot\epsilon$-terms),
${\cal Z}$ is an expression not containing any $\epsilon$, and
\begin{equation}
{\cal G}=\frac{|\epsilon|^{1/2}\, \partial_t \epsilon}{2\epsilon^2}
         \, h^{1/2} \, K \psi^i_i -
         \frac{\partial_t |\epsilon|}{2\epsilon |\epsilon|^{1/2}}
         \, h^{1/2} K^{ij}\psi_{ij}.
\label{z.5}
\end{equation}
If the identification ${\cal F}\equiv 0$ is made, this
reduces to
\begin{equation}
{\cal G}=\frac{|\epsilon|^{1/2}\, \partial_t \epsilon}{2\epsilon^2}\,
          \widetilde{\pi}^{ij}\psi_{ij}.
\label{z.6}
\end{equation}
In this case, apart from the divergence term, the variation of
${\overline{\cal L}}$ fits into the scheme
$G_{\mu\nu}[g]\delta g^{\mu\nu}$, with $G_{\mu\nu}$ being
the full Einstein tensor (including {\it some}
$\dot{\epsilon}$-terms, as is easily obtained from the full
Ricci tensor (\ref{2.7})--(\ref{2.9}) in a zero shift
coordinate system). As a consequence, the ''additional terms''
in (\ref{1.1}) vanish (if the off-shell variations are
constrained by a differentiability condition,
see Section 5), and one recovers the Einstein equations
$R_{\mu\nu}=0$ being equivalent to the variational problem
defined by the Lagrangian density ${\overline{\cal L}}$.
\medskip

A further alternative, that will not be persued here, is to
define the volume element using the complex valued
expression $(-g)^{1/2}$ instead of
$|g|^{1/2}$. This is closer to the spirit of the
''Wick rotation'' in quantum field theory, and to what one
does in Euclidean quantum gravity, and it produces
an additional $i$ in the Euclidean sector. Although
an action generated along these lines may be well suited as
an exponent in a path integral, it does not define a well-posed
classical variational problem.
Its advantage is analyticity rather than reality.
Maybe some of the misunderstandings
in the topic of signature change are due to the appearance
of this (formal) alternative.
\medskip

When applied to
a metric which is in Gaussian form, the expression for
$\delta \cal L$
simplifies considerably. Let us note in particular that
all contributions originating from the variation of the shift
vector (\ref{2.12}) vanish except for the standard Einstein
equation term ${\cal E}^i$ and a spatial divergence. This
is important, because it suspends us from dealing
with non-zero shift metrics. From now on we will ignore
variations with respect to $N_i$.
\medskip

Having noticed this simplification, the variational
principle is defined as follows:
Consider a metric in Gaussian form
\begin{equation}
ds^2=-\epsilon(t) dt^2 + h_{ij}(t,{\bf x})dx^i dx^j
\label{2.14}
\end{equation}
and try to solve the equation $\delta S=0$ for variations
{\it off} this given metric. Part of the information contained
in this equation are the single-signature Einstein equations
off the hypersurface $\Sigma$. We assume these to be already
satisfied (hence ${\cal E}={\cal E}^i={\cal E}^{ij}=0$
and ${\overline{\cal E}}=0$ in the alternative
approach to the volume element problem).
The remaining information in $\delta S$=0 is only relevant
at $\Sigma$, and thus is either inconsistent or leads to
junction conditions, expressed in terms of Gaussian
coordinates. The minimal {\it \`a priori} requirement for
the functions $h_{ij}$ in (\ref{2.14}) is that they are
continuous in $t$ across $\Sigma$
and constitute a strictly positive matrix.
The independent variations to be considered from now on
are $\delta N$ and $\delta h_{ij}\equiv \psi_{ij}$. We assume
these to be continuous across $\Sigma$ as well.
(As will turn out, the variations induced by $\delta N$ to
not produce any new information, once the implications of
$\delta S/\delta h_{ij}=0$ are exploited).
The structure of (\ref{2.14})
implies that $\delta h_{ij}$ provides a variation {\it within}
the class of Gaussian form metrics. As a consequence,
$h_{ij} + \delta h_{ij}$ defines a prototype for an off-shell
metric and has to be continuous across $\Sigma$ as well.
Note that this class of off-shell metrics (within which
all variations are performed) -- although Gaussian
as (\ref{2.14}) -- is conceptually different from
the class of solutions (on-shell metrics).
So far, we have said that off-shell metrics must
have continuous $h_{ij}$.
As we will see, most candidate actions require an additional
({\it \`a priori})
restriction of the class of off-shell metrics (and thus of off-shell
variations $\delta h_{ij}$) in order to define a Lagrangian
model at all. The restrictions that will appear are:\\
{\bf (i)} the $h_{ij}$ are $C^1$ across $\Sigma$
     (i.e. $\partial_t h_{ij}$ exists on $\Sigma$), or\\
{\bf (ii)} the $h_{ij}$ are $C^1$ across $\Sigma$, and
    $\partial_t h_{ij}=0$ there.\\
To what extent the class (ii) can be phrased ''off-shell''
is a matter of taste. In all cases admitting solutions to the
variational problem at all, these solutions
(the on-shell metrics) will
(apart from the regular Einstein-equations off $\Sigma$)
satisfy the condition (i) (which is identical to the junction
conditions for {\it weak} signature change) or,
in addition, (ii) (which is identical to the junction conditions
for {\it strong} signature change). We have
laid some emphasis on these issues because the usual notion
of ''arbitrary variation'' is too rough for signature change.
\medskip

\section{Candidate actions}
\setcounter{equation}{0}

The setting for the action integrals is as follows: Consider
a single signature change at $\Sigma\equiv \Sigma_0$.
Assume $\Sigma$ to be spacelike with respect to its
Lorentzian side.
Denote the total manifold by $\cal M$, and its domains
($t<0$, $t>0$) by (${\cal M}_{-}$, ${\cal M}_{+}$).
In the case of global complications, redefine $\cal M$ to be a
sufficiently close neighbourhood of $\Sigma$. Let
$\epsilon_{\pm}\equiv\epsilon[g_{\pm}]$ denote the metric
signature in these domains ($\epsilon[g]=$
$\epsilon_{-}\Theta(-t) +$ $\epsilon_{+}\Theta(t)$ being
the discontinuous ''covariant'' signature expression
for the total manifold $\cal M$; its precise value
at $\Sigma$ will never be important).
In the case of a true
signature change we have $\epsilon_{+}=-\epsilon_{-}$
($=1$ or $-1$). However, for comparison we keep the
possibility $\epsilon_{+}=\epsilon_{-}$ ($=1$ or $-1$). From
now on, sub- or superscripts $\pm$ will distinguish
quantities due to ${\cal M}_{\pm}$.
\medskip

The singular actions are (formally) defined as
\begin{equation}
S_1 = \int_{\cal M} d^n x\,{\cal L}
\label{2.s1}
\end{equation}
and
\begin{equation}
S_2 = \int_{\cal M} d^n x\,\epsilon[g]\,{\cal L}.
\label{s.s2}
\end{equation}
As explained in the previous Section,
$S_1$ provides one interpretational version of the
Einstein-Hilbert action. The other version is defined by
(using ${\overline{\cal L}}$ from (\ref{z.2}) with
${\cal F}\equiv 0$)
\begin{equation}
S_3 = \int_{\cal M} d^n x\,{\overline{\cal L}}.
\label{y.1}
\end{equation}
For completeness, we also introduce the analogue of $S_2$ within
this alternative scheme
\begin{equation}
S_4 = \int_{\cal M} d^n x\,\epsilon[g]\, {\overline{\cal L}}.
\label{y.2}
\end{equation}
We should add here that ony might hope the principle of
''general covariance'' to single out $S_1$ or $S_3$ as the
''correct'' action generalizing the standard $S_{EH}$.
At least, in ordinary general relativity, one is of
course not free to modify the volume element by will.
However, in a theory admitting a change of signature, an
additional coordinate-independent object arises, namely the
hypersurface $\Sigma$ at which the change occurs, together
with its geometric (intrinsic and extrinsic) structur.
Accordingly, the difference $S_3 - S_1$ is just a surface integral
over $\Sigma$. Inserting $|\epsilon_{\pm}|=1$, we get
\begin{equation}
S_3 = S_1 + (\epsilon_{+}-\epsilon_{-})
       \int_{\Sigma} d^{n-1} x\,h^{1/2}\,K ,
\label{diff}
\end{equation}
hence none of the two
actions can be excluded by covariance arguments. As we have
already stated in Section 2, the variational principle
defined by $S_3$ is equivalent to the full (distributional)
Einstein equations. Hence, equation (\ref{diff}) shows that
the straightforward generalization of $S_{EH}$,
namely $S_1$, must be modified by a hypersurface integral
in order to reproduce the full Einstein equations from a
Lagrangian model.
\medskip

The first two regular actions result, as announced, from
breaking the singular ones into
\begin{equation}
S_5 = - \,\epsilon[g_{-}] \int_{{\cal M}_{-}}
           d^{n}x\,|g_{-}|^{1/2} R[g_{-}] +
      \epsilon[g_{+}] \int_{{\cal M}_{+}} d^{n}x\,
           |g_{+}|^{1/2} R[g_{+}]
\label{2.s3}
\end{equation}
and
\begin{equation}
S_6 = \epsilon[g_{-}] \int_{{\cal M}_{-}} d^{n}x\,
        |g_{-}|^{1/2} R[g_{-}]+
      \epsilon[g_{+}] \int_{{\cal M}_{+}} d^{n}x\,
        |g_{+}|^{1/2} R[g_{+}].
\label{2.s4}
\end{equation}
It proves useful to define
\begin{equation}
S_{\pm} = \epsilon[g_{\pm}] \int_{{\cal M}_{\pm}}
       d^{n}x\,|g_{\pm}|^{1/2} R[g_{\pm}],
\label{2.spm}
\end{equation}
then $S_5=-S_{-}+S_{+}$ and $S_6=S_{-}+S_{+}$.
\medskip

The remaining two regular actions are constructed by subtracing the
$\partial_t$-term from (\ref{2.5}), hence they consist of
integrations over the Lagrangian densities
\begin{equation}
{\cal L}_{\pm} =
\epsilon_{\pm} N_{\pm}\,h_{\pm}^{1/2}\,R[g_{\pm}]
   + 2 \,\partial_t (\,h_{\pm}^{1/2}\, K_{\pm}).
\label{2.lpm}
\end{equation}
In view of the sign structure appearing here, a slight redefinition
is in place. So far, the extrinsic curvature $K_{ij}$ has been
defined  with respect to the normal vector $n^\mu$
which is (at $\Sigma$) {\it out}ward directed with respect to
${\cal M}_{-}$
and {\it in}ward directed with respect to ${\cal M}_{+}$.
In order to have a notation at hand that relies on the
intrinsic properties of a domain alone, we define
${\cal K}_{ij}$ to be the extrinsic curvature with respect to
the {\it in}ward directed normal of each domain,
and ${\cal K}={\cal K}^i_i\equiv $ $h^{ij}{\cal K}_{ij}$.
Hence, ${\cal K}_{ij}^{\pm}=\pm K_{ij}^{\pm}$ and
(due to the continuity of $h_{ij}$) ${\cal K}_{\pm}=\pm K_{\pm}$.
As a consequence, the integration over (\ref{2.lpm})
results into ${\cal S}({\cal M}_{\pm})$, where for any
single-signature domain $\cal N$ we have defined
\begin{equation}
{\cal S}({\cal N}) =
  \epsilon[g] \int_{\cal N} d^n x\, |g|^{1/2}\,R[g]
 - 2 \int_{\partial {\cal N}} d^{n-1} x\,|h|^{1/2}\,{\cal K}.
\label{2.s}
\end{equation}
This is a natural construction, and it is very close to the action
used in the path integral formulations for quantum gravity
\cite{GibbonsHawking},\cite{Hawking1}
and quantum cosmology
\cite{HartleHawking},\cite{Hawking2}.
Note that $\partial {\cal M}_{\pm} = \Sigma$ for both
domains, but with a different orientation. This is equivalent
to the fact that the $\partial_t$-term in (\ref{2.lpm}),
when integrated over $t$, produces values on $\Sigma$, but
with an additional $\mp$ sign. Since we are only interested
in the junction conditions at $\Sigma$, we neglect any
contribution
from other boundaries of ${\cal M}_{\pm}$ than $\Sigma$.
The two remaining candidate actions are thus
\begin{equation}
S_7= - {\cal S} ({\cal M}_{-}) + {\cal S} ({\cal M}_{+})
\label{2.s5}
\end{equation}
and
\begin{equation}
S_8=   {\cal S} ({\cal M}_{-}) + {\cal S} ({\cal M}_{+}).
\label{2.s6}
\end{equation}
\medskip

By looking at (\ref{2.5}) we see that the $\epsilon$-prefactors
of $S_{\pm}$ and ${\cal S}({\cal M}_{\pm})$
cancel those in front of the kinetic term in ${\cal L}$. For
$\epsilon_{+} =$ $ - \epsilon_{-} =1$, the actions
$S_5$,\dots $S_8$ take the form as described in the
introduction. However, had we assumed this standard
sign choice from the outset, we would not have noticed
the signature pattern provided by the various $\epsilon$'s
that makes $\epsilon[g]R[g]$ a particularly nice alternative
to $R[g]$. We are now ready to discuss the eight candidate
actions. The properties of the singular actions rely on the explicit
$\dot{\epsilon}$-contributions appearing in the variational formulae
of Section 2 (these terms do not even show up in the regular
actions). The properties of the regular actions
are based on the partial $t$-derivative terms that
are, from the singular action point of view, just divergences
that ought to be omitted.
\medskip

\section{Singular actions $S_1$ and $S_2$}
\setcounter{equation}{0}

The variation $\delta S_1$ is obtained by integrating
(\ref{2.11}) and (\ref{2.13}) over the total manifold.
Assuming the metric considered
to be of the Gaussian form, inserting the regular
part of the Einstein equations ${\cal E}={\cal E}^{ij}=0$
and throwing away the boundary terms, we note that a
$\dot{\psi}_{ij}$-term has still survived. Integrating by
parts (in the sense of distributions), this can be
removed, leading to a term proportional to
$\ddot{\epsilon} h^{1/2}\,h^{ij}\delta h_{ij}/\epsilon^2$. In the
case of a true signature change, $\ddot\epsilon$ is of
the type $\delta'(t)$. If arbitrary off-shell variations
$\delta h_{ij}$ that are continuous (or $C^1$) across
$\Sigma$ are admitted, an immediate consequence is
$h^{ij}=0$, which is unacceptable. Hence, the
standard Einstein-Hilbert action $S_1$
(when interpreted as the integral over the product of
$|g|^{1/2}$ with $R[g]$ as two quantities by their own)
does {\it not} define
a sensible variational principle for signature change. One
may of course restrict the class of off-shell metrics to
$h_{ij}$ being $C^1$ and satisfying $\partial_t h_{ij}=0$
at $\Sigma$ (condition (ii) in Section 2). As a
consequence, $K_{ij}=0$, and only variations with
$\dot{\psi}_{ij}\equiv $ $\partial_t \delta h_{ij}=0$
are allowed. This gives $\delta S_1=0$, and the junction
conditions of {\it strong} signature change, but
the junction conditions have been assumed rather than
derived. Also, the notion of off-shell metrics having
to satisfy such a strong condition is not satisfactory
at all.
\medskip

The situation is even worse with $S_2$. Multiplying
(\ref{2.11}) and (\ref{2.13}) by $\epsilon$ and integrating
by parts, we obtain the relevant contribution
\begin{equation}
\frac{\dot\epsilon}{2\epsilon}\,h^{1/2}\,
    \bigg( K\delta N+ K \psi^i_i - 2 K^{ij}\psi_{ij} -
          h^{ij}\dot{\psi}_{ij} \bigg)
\label{3.1}
\end{equation}
to be integrated over. Here $\dot\epsilon/\epsilon$ is not
even a well-defined distribution (it is of the type
${\rm sgn}(t) \delta(t)$). The only possibility to make
the parenthesis vanish is to restrict the class of off-shell
metrics to satisfy the {\it strong} junction conditions
(condition (ii) of Section 2) as above, which is against the
spirit of a Lagrangian formalism as well.
Alternatively, one may imagine $\epsilon(t)$
to arise as a limit of antisymmetric functions
$\epsilon_{\rm smooth}(t)$. Then formally,
$\dot{\epsilon}/\epsilon$ is zero when
multiplied by continuous functions. As a
consequence, the $\dot{\psi}_{ij}$-term drops out
(its continuity may be assumed), and the rest
in the parenthesis of (\ref{3.1}) is continuous if $K_{ij}$
is (which implies {\it weak} signature change). This
procedure suffers from the ignorance of
$|g|^{1/2}\equiv$ $N\,h^{1/2}$ against the contributions
of $\epsilon_{\rm smooth}(t)$ {\it before} the limit is
taken. However, it may be considered as an
{\it ad hoc} regularization scheme for $S_2$.
\medskip

Thus, the perspectives for using $S_1$ and $S_2$ as actions are
somewhat unpleasant and limited.
\medskip

\section{Singular actions $S_3$ and $S_4$}
\setcounter{equation}{0}

As already stated in Section 2, the variational principle
$\delta S_3=0$ generates the full Einstein equations.
This is the
Lagrangian version of the usual point of view that
assumes the Einstein equations as first principles.
The variational formula (\ref{z.4}) with ${\cal F}\equiv 0$ and
(\ref{z.6}) immediately implies that,
upon imposing $\partial_t h_{ij}$ to be continuous for the
class of off-shell metrics (in order to make $S_3$ well-defined;
this is the condition (i) of Section 2),
the solutions must satisfy $\widetilde{\pi}^{ij}=0$ at $\Sigma$.
For $n \neq 2$ this is equivalent to $K_{ij}=0$, hence
{\it strong} signature change. Thus, $S_3$ defines
a Lagrangian model, as long as one accepts the differentiability
condition for the off-shell metrics.
\medskip

The candidate action $S_4$ provides similar problems as
$S_2$. Instead of expressions like
$\dot{\epsilon} |\epsilon|^{1/2}/\epsilon^2$
(which appear in $\delta S_3$ and can -- after setting
$|\epsilon|=$ $\epsilon^2=1$ in the end -- be interpreted
as ordinary $\delta$-distributions), we are now faced with
$\dot{\epsilon} |\epsilon|^{1/2}/\epsilon$, which is much
less well-defined. The possible (although not very appealing)
procedures one may perform are analogous to those
mentioned in the discussion of $S_2$ in the previous Section.
Hence, we skip all details and just summarize
that $S_4$ is ruled out along the same lines as
$S_2$.

\section{Regular actions $S_5$ and $S_6$}
\setcounter{equation}{0}

We write down the variations of $S_{\pm}$ off a metric
in Gaussian form. Using (\ref{2.11}) and (\ref{2.13}),
we get
\begin{equation}
\delta S_{\pm} =
\mp \int_{\Sigma} d^{n-1} x \,h^{1/2}\,
   \bigg( 2 K^{\pm} \delta N + K^{ij}_{\pm}\psi_{ij}
        + h^{ij} \dot{\psi}_{ij}^{\pm}  \bigg).
\label{4.1}
\end{equation}
Note that the $\epsilon$'s have cancelled due to the prefactors.
Considering now the candidate action $S_5=-S_{-}+S_{+}$, we
encounter the contribution
$h^{ij}(\dot{\psi}_{ij}^{+} + \dot{\psi}_{ij}^{-})$ which has
to vanish in order to allow for solutions of $\delta S_5=0$.
Again, a restriction of the class of off-shell metrics
is necessary. The only natural way to do this is to impose the
{\it strong} junction conditions (condition (ii) from Section 2)
in order to enforce $\dot{\psi}_{ij}^{\pm}=0$. Thus, $S_5$ does
not define a sensible model, along with $S_1$, $S_2$ and $S_4$.
\medskip

The situation first changes with $S_6= S_{-}+S_{+}$. The unpleasant
$\dot{\psi}_{ij}^{\pm}$-terms appear with a relative minus
sign, and we are left with
\begin{equation}
\delta S_6 =
\int_{\Sigma} d^{n-1} x \,h^{1/2}\,
   \bigg( 2 (K^{-}-K^{+}) \delta N
       +   ( K^{ij}_{-} - K^{ij}_{+} )\psi_{ij}
         + h^{ij} (\dot{\psi}_{ij}^{-}
      - \dot{\psi}_{ij}^{+}) \bigg).
\label{4.2}
\end{equation}
Restricting the class of off-shell metrics to those
for which $h_{ij}$ is $C^1$ across $\Sigma$
(condition (i) of Section 2), we get
$\dot{\psi}_{ij}^{-}-\dot{\psi}_{ij}^{+}=0$. The remaining
variational problem leads to $K^{ij}_{-}=K^{ij}_{+}$,
hence {\it weak} signature change. We conclude that $S_6$ defines
a sensible model, once the differentiability condition
defining the class of off-shell metrics is accepted.
\medskip

Note that all derivations given in this Section apply for
the single-signature case $\epsilon_{+}=\epsilon_{-}$ as well.
For $S_6$ this becomes what we have called the
''additivity property''
in the introduction: one may (assuming the off-shell-$h_{ij}$ to
be $C^1$ across $\Sigma$) derive the standard Einstein
equations by breaking the Einstein-Hilbert action into
a sum of two integrals. This may be understood as a naturality
argument in favour of $\epsilon[g]R[g]$.
\medskip

\section{Regular actions $S_7$ and $S_8$}
\setcounter{equation}{0}

These actions provide the least complications, since the
''time''-derivative part of (\ref{2.5}) is missing. One
easily finds
\begin{equation}
\delta {\cal S}({\cal M}_{\pm}) =
    \mp \int_{\Sigma} d^{n-1} x\, h^{1/2}\,
       (K_{\pm} h^{ij} - K_{\pm}^{ij})\psi_{ij}
   \equiv \mp \int_{\Sigma} d^{n-1} x\,
            \widetilde{\pi}_{\pm}^{ij} \psi_{ij}.
\label{5.1}
\end{equation}
For
$S_7= - {\cal S} ({\cal M}_{-}) + {\cal S} ({\cal M}_{+})$,
this gives an integrand
$(\widetilde{\pi}^{ij}_{+} + \widetilde{\pi}^{ij}_{-})\psi_{ij}$.
Hence, when the off-shell metrics are restricted to
continuous $\partial_t h_{ij}$ (condition (i) from Section 2),
we have
$\widetilde{\pi}^{ij}_{+} = \widetilde{\pi}^{ij}_{-}$. As a
consequence, we find $\widetilde{\pi}^{ij}=0$ at $\Sigma$
as the solution of the variational problem. For $n \neq 2$
this is equivalent to $K_{ij}=0$ at $\Sigma$. Hence,
$S_7$ provides a Lagrangian model allowing for
{\it strong} signature change, once the class of off-shell
metrics is characterized by the above differentiability
condition.
\medskip

However, there is another possibility to let $S_7$ define
a variational problem:
If one is willing to give up the continuity of $K_{ij}$
across $\Sigma$
(and hence keeps the class of admissable off-shell metrics
unconstrained)
the action $S_7$ leads
to $K_{ij}^{-}+K_{ij}^{+}=0$. This predicts a kink in the
induced metric across $\Sigma$ and could in principle be
considered as a further ({\it super-weak} or
''anticontinuous'') junction condition
by its own \cite{Drayppp}.
\medskip

The variation of the last action
$S_8=   {\cal S} ({\cal M}_{-}) + {\cal S} ({\cal M}_{+})$ leads
to the integrand
$(\widetilde{\pi}^{ij}_{+} - \widetilde{\pi}^{ij}_{-})\psi_{ij}$.
Hence, without restricting the class of off-shell metrics
(apart from $h_{ij}$ being continuous), this immediately leads to
the solution $\widetilde{\pi}^{ij}_{+} = \widetilde{\pi}^{ij}_{-}$
of the variational problem. If $n \neq 2$, this is
equivalent to $K_{ij}^{+} = K_{ij}^{-}$. In this way, $S_8$
defines a perfectly well-posed variational principle allowing
for {\it weak} signature change. It is the only candidate
action that requires no differentiability assumption for the
class of off-shell metrics (but rather {\it predicts}
$\partial_t h_{ij}$ to be continuous). It is thus closer
to the spirit of a Lagrangian formalism than all other candidates.
Moreover, for $\epsilon_{+}=\epsilon_{-}$, it shows
the ''additivity property'' (and thus leads to the
single-signature Einstein equations straightforwardly).
In this sense, {\it weak} signature change
seems favoured from the Lagrangian point of view.
\medskip

\section{Variations including $\delta\Sigma$}
\setcounter{equation}{0}

So far we have fixed the hypersurface $\Sigma$ of signature change.
However, a {\it full} action principle requires the structure
(${\cal M}_{-}$, $g_{-}$, ${\cal M}_{+}$, $g_{+}$) to
be interpreted as a {\it single} configuration with
respect to which the action is varied.
\medskip

All of our candidate actions may be written as a sum of
a volume-integral over ${\cal M}_{\pm}$ and a surface-integral
over $\Sigma$.
Start from an on-shell metric (i.e. satisfying the
field equations and junction conditions) and allow $\Sigma$ to be
deformed infinitesimally into $\Sigma^{\rm new}$. Choose an
(off-shell) metric that changes sign at $\Sigma^{\rm new}$.
The variation
of the volume-integral part typically yields --
apart from contributions that vanish on account of the
field equations -- expressions of the form
\begin{equation}
\bigg(\int_{{\cal M}_{+}^{\rm new}} -\int_{{\cal M}_{+}}
        \bigg) d^n x\,{\cal L}_{+}
\label{8.var1}
\end{equation}
and a similar term for ${\cal M}_{-}$.
Next we use the fact that the volume part of the Lagrangian
densities vanishes
at the hypersurface of signature change (see equation
(\ref{q3}) below for a justification --
this carries over to the more general case
when matter fields are included: the volume part of the
Lagrangian density
will not necessarily vanish on-shell in general, but it {\it does}
so at the hypersurfaces of signature change).
Hence we may, to first order,
insert ${\cal L}_{+}=0$ and find zero variation due to the
volume-integral contributions. The contributions to
$\delta S$ that represent surface-integrals turn out to
vanish partially due to fact that the junction conditions are
satisfied by the old metric at the old
hypersurface. However, for those Lagrangians which
contain true surface contributions (the integrand being
a multiple of $K$ in each case) some terms of the type
\begin{equation}
\bigg(\int_{\Sigma^{\rm new}} - \int_{\Sigma}
        \bigg) d^{n-1} x\, h^{1/2}\, K
\label{8.var2}
\end{equation}
remain. In the case of $S_7$ and $S_8$, these are manifestly
absent (as is the case for $S_3$ and -- in one of the
two regularization schemes -- for $S_4$).
Thus we find $\delta^{\rm tot} S_{3,4,7,8}=0$.
(In the next section we will see that $S_7$ and $S_8$
are essentially equivalent to $S_3$ and $S_4$). In the
other cases, the condition $\delta S=0$ puts further
constraints on the class of admissable off-shell metrics
or, if one is not willing to implement these, restricts
the set of solutions even further. (To be a bit more specific
we note that in the vaccum case the variation might still
vanish in some cases because the field equations imply
$\partial_t K_{\pm}=0$ at $\Sigma$, but when a
cosmological constant is included, this generalizes to
$\partial_t K_{\pm}=$ $-2\epsilon_{\pm} N \Lambda/(n-2)$, where
$n=\delta^\mu_\mu$).
\medskip

Summarizing, we found that for the most interesting actions
$S_{3,4,7,8}$ the variation of the
signature change hypersurface $\Sigma$ does not
provide any new information.
\medskip

\section{$S_7$ and $S_8$ revisited}
\setcounter{equation}{0}

Putting our previous results together, the two candidate actions
$S_7$ and $S_8$
provide the most reasonable actions for {\it weak}
and {\it strong} signature change, respectively.
(We leave apart
the possibility of obtaining an ''anticontinuous''
junction condition from $S_7$, as was mentioned in
Section 7).
So far, these
action functionals have been defined in a
''covariant'' (i.e. coordinate-independent) way (see
(\ref{2.s5}) and (\ref{2.s6})).
However, it is instructive to write down the expressions
for $S_7$ and $S_8$ in terms of a coordinate system in which
the metric is given by (\ref{2.1}), with
$N$, $N_i$ and $h_{ij}$ being continuous and $\Sigma$ described
by the equation $t=0$.
The extrinsic curvature $K_{ij}$, as given by (\ref{2.2}),
plays (apart from the lapse factor and the shift
contribution) the role of a ''velocity'', with $h_{ij}$
being the basic dynamical variable.
\medskip

Joining the two integrals of (\ref{2.s5}) into a single one,
we may write
\begin{equation}
S_7 = \int_{\cal M} d^n x\,N\,h^{1/2}\,\bigg(
        {\widetilde{\epsilon}}_1\, \Big(K^{ij}K_{ij}-K^2 \Big)
         + {\widetilde{\epsilon}}_2\, R[h] \bigg),
\label{q1}
\end{equation}
where, ${\widetilde{\epsilon}}_1(t)={\rm sgn}(t)$ and
${\widetilde{\epsilon}}_2(t)=-\epsilon_{-}\Theta(-t) + $
$\epsilon_{+}\Theta(t)$. If $\epsilon_{+}$ and $\epsilon_{-}$
are negative to each other, ${\widetilde{\epsilon}}_2(t)$
is constant. In particular, for $\epsilon_{+}=-\epsilon_{-}=1$,
we get ${\widetilde{\epsilon}}_2(t)=1$.
As a by-product we note that,
applying (\ref{y.1}) for (\ref{z.2}) with ${\cal F}\equiv 0$,
and inserting $|\epsilon|=1$ (which is allowed in this place,
due to our construction), we find the same expression,
hence $S_3= \pm\, S_7$ if $\epsilon_{+}=-\epsilon_{-}=1$.
The non-standard definition of
$|g|^{1/2} R[g]$ just generates the same action as the
two-domain construction for $S_7$.
Martin also arrives at an identical expression (in a
slightly more heuristic way) \cite{Martin}.
Note that $K_{ij}$ contains a factor $1/N$, so that
(when expressed in terms of the ''velocities''
$\dot{h}_{ij}$), the overall structure of this expression
is $\,\,$''(1/N) kinetic energy -- N potential energy''. The
variation with respect to $N$ generates the Hamiltonian constraint
equation (although in Lagrangian form, see equation
(\ref{q3}) below). The variation
with respect to $h_{ij}$ (generating the time evolution
equations) produces the $\delta$-function terms contained in
the full Einstein equations, hence the junction conditions
for {\it strong} signature change.
\medskip

One may bring $S_8$ into an analogous form by joining the two
integrals of (\ref{2.s6}) into a single one. The reason for this
is that the construction of $S_8$ is just the covariant formulation
of the recipe to omit the $\partial_t$ and
$\dot{\epsilon}$-terms in $\epsilon {\cal L}$, with
${\cal L}$ given by (\ref{2.5}). Hence,
\begin{equation}
S_8 = \int_{\cal M} d^n x\,N\,h^{1/2}\,\bigg(
        K^{ij}K_{ij}-K^2
                  + \epsilon \, R[h] \bigg).
\label{q2}
\end{equation}
The factor $\epsilon$ arises here straightforwardly.
(Let us note in parentheses that the regularization scheme
for $S_4$ that puts $\dot{\epsilon}|\epsilon|^{1/2}/\epsilon$
effectively equal to zero yields the same expression,
hence in this sense $S_4=S_8$. A similar argument shows that,
in the same sense, $S_2=S_6$).
Again, the overall structure is $\,\,$
''(1/N) kinetic energy -- N potential energy'', but with the signs
(and sign changing terms) distributed differently as compared
to (\ref{q1}). In contrast to $S_7$, the variation with respect
to $h_{ij}$ does not produce any $\delta$-functions.
The field equations just contain $\epsilon$ in a purely algebraic
way, together with
$\dot{K}^{ij}$ as ''accelerations'', which are thus
predicted to undergo a finite jump. Hence, upon
integration, the ''velocitiy''-type quantities $K_{ij}$
must be well-defined across $\Sigma$, which is identical to the
junction conditions for {\it weak} signature change.
Note also that the sign structure appearing in these two
actions corresponds to that of the toy model Lagrangians
(\ref{1.2}) and (\ref{1.3}) given in the introduction.
\medskip

The constraint equation, $\delta S/\delta N=0$, is the same
for both actions if $\epsilon_{+}=-\epsilon_{-}=1$,
and reads
\begin{equation}
K^{ij}K_{ij}-K^2 = \epsilon\, R[h].
\label{q3}
\end{equation}
Thus, whenever $\epsilon$ changes sign, the continuity
of the extrinsic curvature (which is a consquence in both
models) implies that at signature change all solutions must
satisfy
\begin{equation}
K^{ij}K_{ij}-K^2 = R[h] = 0.
\label{q4}
\end{equation}
In the model $S_7$, the additional condition is $K_{ij}=0$.
In any case, the Lagrangian density vanishes at $\Sigma$
(this was used in a somewhat different picture in the previous
Section when $\delta \Sigma$ was carried out).
\medskip

In a Friedmann-Robertson-Walker model with $N=1$ and scale
factor $a(t)$, the kinetic terms are given by
$K^{ij}K_{ij}-K^2=$ $-(n-1)(n-2)\dot{a}^2/a^2$,
which is non-positive (contrary to the generic case). As a
consequence,
(\ref{q4}) implies $\dot{a}=0$, hence $K_{ij}$=0. Thus, in this
highly symmetric case, {\it weak} signature change
automatically implies the {\it strong} junction conditions.
This remains true if a cosmological constant is included, but
already fails for the case of an additional scalar field,
although Hayward \cite{Hayward3} seems to claim the contrary
in his criticism of Ellis {\it et al}
\cite{EllisSumeruketal}: his
equations (23) -- version December 1994 -- {\it do} admit
true {\it weak} signature changing solutions in the sense
we use this notion. In terms of $t$ they have
discontinuous but bounded $\partial_{t t} a$, but
when expressed in terms of a
''continuous'' approach time coordinate $\tau$ (cf. equation
\ref{ccc}), these solutions display singular second derivatives,
$\partial_{\tau\tau} a\sim \epsilon |\tau-\tau_0|^{-1/2}$, which is
the reason for Hayward not to talk about them at all. This is an
example of different results based on different first principles.
\medskip

We would like to add a remark concerning a possible
restriction of the signature change models.
This requires to add a cosmological constant contribution
to all actions. The inclusion of a cosmological constant
is achieved by replacing
\begin{equation}
R[g]\rightarrow R[g]-2\Lambda, \qquad
R[h]\rightarrow R[h]-2\Lambda.
\label{q5}
\end{equation}
The proposal we have in mind starts from the fact that,
so far, we are not able to
be predict whether a signature change will actually
occur when it is possible (e.g. when $R[h]$ becomes zero
in the $S_8$-model; cf. the remarks made in the introduction).
This is reflected by the fact that the signature dependent
factors $\epsilon$ and ${\widetilde{\epsilon}}_2$ in
the expressions (\ref{q1}) and (\ref{q2}) above are {\it not}
functionals of the dynamical variables $h_{ij}$. Their
dynamical status is actually rather obscure in a Lagrangian
(or Hamiltonian) formalism, and when trying to quantize
these models, one wonders how to treat them
(cf. Ref. \cite{Martin}).
Classically, from (\ref{q3}), one reads off that $\epsilon$
is the sign of $(K^{ij}K_{ij}-K^2)R[h]$, and inserting
this into $S_8$ reveals its involved structure.
Moreover, the coordinate system used here restricts any
possible signature change to occur at a hypersurface of
constant $t$. Another coordinate system would produce
signature changes along different hypersurfaces.
However, inspired by quantum cosmology, one may think
about fixing a cosmologically preferred coordinate system
(e.g. by requiring the shift vector $N_i$ to vanish and $N$ to
depend only on $t$). Such a choice is dynamically
consistent within the framework of minisuperspace
models, where part of the degrees of freedom are frozen and
integrated over. Let us denote the ''potential''
occuring in such models by $2 \Lambda-R[h]$, although $R[h]$
might actually be an integrated version of the Ricci
scalar.
When the universe ''is born'' at small geometries
(which implies $2 \Lambda-R[h] <0$) with Euclidean signature
($\epsilon=-1$), it will eventually ''evolve'' (with
respect to $t$) towards the conditions for a signature
change. An {\it ad hoc} assumption resolving the
non-uniqueness problem for solutions is to postulate that
whenever $2 \Lambda-R[h] <0$ (which is often called the
Euclidean sector in superspace) the signature is Euclidean
($\epsilon=-1$),
and whenever $2 \Lambda-R[h]>0$, the signature is Lorentzian
($\epsilon=1$).
This is (together with the initial condition) just the
statement that a signature change {\it will} occur
whenever it is possible.
Upon inserting
$\epsilon = {\rm sgn} (2\Lambda-R[h])$ into $S_8$,
$\epsilon$ has become a functional of the dynamical
variables. This modification gives a nice expression, namely
\begin{equation}
S_8' = \int_{\cal M} d^n x\,N\,h^{1/2}\,\bigg(
            K^{ij}K_{ij}-K^2 - \Big| 2 \Lambda - R[h] \Big| \bigg).
\label{q6}
\end{equation}
A ten-dimensional version of this action within the framework of a
Friedmann-Robertson-Walker (FRW) minisuperspace model may be found
in Ref. \cite{FE2} (where this type of signature change
is shown to stabilize internal dimensions).
Here, quantization is straightforward
(at least formally), and one obtains a Wheeler-DeWitt equation
\cite{DeWitt} with the ''potential term'' being replaced by
its alsolute value. Similar constructions are of course possible
when matter fields are included in addition (or instead of)
the cosmological constant.
The modification of $S_7$
along the same lines gives a model that is a bit more involved,
and (classically) allows for less signature changes than $S_8$.
The quantization of such a model seems more problematic
(naively, one would obtain the same Wheeler-DeWitt equation
as for $S_8'$), but being quantized does not seem to be the
intention of {\it strong} signature change anyway (as a classical
model of quantum tunneling).
\medskip

Finally, we give the overall structure of $S_7$, $S_8$
and $S_8'$
in the simplest case of a four-dimensional FRW model with
cosmological constant by writing down the Lagrangians
\begin{eqnarray}
{\cal L}_7^{\rm FRW}
&=& - {\rm sgn}(t) \dot{a}^2/N + N ( a^2 - \Lambda a^4 ),
\label{frw7}\\
{\cal L}_8^{\rm FRW}
&=& -  \dot{a}^2/N + N {\rm sgn}(t) ( a^2 - \Lambda a^4 ),
\label{frw8}\\
{\cal L}_8^{\rm FRW ' }
&=& -  \dot{a}^2/N - N | a^2 - \Lambda a^4 |,
\label{frw8prime}\\
\nonumber
\end{eqnarray}
where the scale factor and the lapse function have been
conveniently redefined.
\medskip

\section{Conclusion}
\setcounter{equation}{0}

Our analysis has singled out $S_3$, $S_4$ (in a rather
subtle interpretation), $S_6$, $S_7$ and $S_8$
to define more or less acceptable
models for signature change ($S_3$ and $S_7$ for the {\it strong},
the others for the {\it weak} junction conditions) in any
total space-''time'' dimension $n\geq 3$. Moreover we found that
$S_3$ is essentially identical to $S_7$ (and in some sense
$S_4$ to $S_8$). Straightforward
quantization might be conceptually problematic for $S_6$,
because of the off-shell metric differentiability
conditions,
the variation of the hypersurface $\Sigma$ (Section 8)
being problematic as well.
Hence, one should consider $S_7$ ($S_8$) as the
best model for {\it strong} ({\it weak}) signature change,
fitting into the Lagrangian framework as regular
actions ($S_7$, when defined as $S_3$, as a
singular action as well,
and $S_8$, when defined as
$S_4$, as a singular action in a particular regularization
scheme).
None of these two models can {\it \`a priorily} be excluded. This
establishes both versions of signature change as
Lagrangian models. The regular actions
seem to be better suited for a variational
description of signature change than the singular ones. In
addition we observed that
$S_7$ needs a differentiability condition for the class of
off-shell metrics, whereas $S_8$ defines a perfectly
well-defined variational problem if just a simple continuity
condition is imposed. In this sence, the Lagrangian point of view
slightly favours {\it weak} over {\it strong} signature change.
(In additon, as was mentioned in Section 7, the relaxation
of the conditions on the class of off-shell metrics for
$S_7$ leads to an ''anti-continuous''
junction condition as a third version of signature change.
It predicts a kink in $h_{ij}$, hence
a non-differentiable induced metric,
but cannot be excluded either ({\it \`a priorily}) within the
regular actions approach).
\medskip

The way we proceeded is something like an ''inverse philosophy''
as compared to the usual approaches. The models associated with
{\it strong} signature change require more severe restrictions
on the off-shell metrics than the {\it weak} models in order
to get things well-defined. Normally, one would interpret these
restrictions (in particular $K_{ij}=0$) as an argument in
favour of {\it strong} signature change. However, in a
Lagrangian framework one is interested in a {\it large}
class of off-shell metrics -- which favours $S_8$ against $S_7$.
This seems to correspond with the different aims pursued
with the two models. {\it Strong} signature change, when
envisaged as a genuinely classical model, might not
{\it need} a Lagrangian formulation, whereas {\it weak}
signature change might be quantized, the goal being an
alternative to standard quantum cosmology.
\medskip

The full power of $S_8$ becomes visible if it is rephrased
as follows: Divide $\cal M$ into a finite (or sufficiently nice
discrete) set of domains ${\cal M}_J$ inside which the
metric has constant signature (and such that
all boundaries $\partial {\cal M}_J$ are spacelike whenever
this is well-defined). Moreover, let the corresponding
induced metrics at the boundaries be continuous. Define
the action to be
\begin{equation}
S =\sum_{J} {\cal S}({\cal M}_J).
\label{7.1}
\end{equation}
Assuming local infinitesimal variations of this situation,
the condition $\delta S=0$ produces {\it weak} signature
change across those boundaries $\partial {\cal M}_J$ at which
$\epsilon[g]$ is discontinuous, and the standard (Lorentzian
or Euclidean) Einstein equations everywhere else. Thus,
in this framework,
{\it weak} signature change appears
at an equal footing with the ''additivity property''
of the Einstein-Hilbert action (with boundary term removed),
and hence seems to be a natural generalization thereof.
This is in turn linked with the conceptual advantages of the
regular actions as compared to the singular ones. Moreover, these
issues seem to relate to the original motivation for subtracting
a boundary term from the Einstein-Hilbert action in
quantum gravity \cite{GibbonsHawking}
\medskip

Can one draw a simple conclusion from all these details? In my
opinion it is the following: Writing down the integrand of the
Einstein-Hilbert action, one encounters distributional and even
worse terms
(i.e. terms involving $\dot{\epsilon}$ in our language).
Among all attempts we tested, exactly those who effectively
remove these terms led to an acceptable variational principle.
As a consequence, one could think about giving up all
''distributional'' interpretations of what happens at
$\Sigma$, and to adopt the idea of regular actions, which are
perfectly well-defined within each domain. Once this is done,
it appears as a viable alternative to reverse the sign in the
Euclidean domains (which, in addition, improves the
well-posedness of the action and the properties of
the admissible off-shell variations). In this way,
both {\it strong} and {\it weak} signature change emerge as
Lagrangian models.
\medskip

Summarizing, let us state that the Lagrangian approach clearly
reveals the existence of (essentially two) different generalizations
of general relativity allowing for a classical change
of metric signature, none of them being {\it \`a priori}
the ''correct'' one.
This justifies our convention to talk about {\it weak} and
{\it strong} signature change as two phenomena, appearing
within different models, which are based on
different first principles. Further
{\it pros and cons} are of course still possible and can be based
on arguments that concern physical predictions, mathematical or
physical richness, naturality or even aestetics.
\\
\\
\\
{\Large {\bf Note added}}
\medskip

Due to discussions that took place after the original
manuscript was prepared, I would like to add a comment on
the singular actions aproach.

As was stated in Section 2, the spacial object $K_{ij}$
coincides with the (invariantly defined) extrinsic curvarure
only if $|\epsilon|=1$. However, in part of the derivations
we have treated $\epsilon$ as if it were an arbitrary
function (and sometimes called this procedure
a ''regularization'') -- with the exception that $K_{ij}$
is always kept as a quantity without further reference
to $\epsilon$.
One may however replace at any stage of the ''regularization''
procedures $K_{ij} = |\epsilon|^{1/2}\,K_{ij}^{\rm true}$ and
keep $K_{ij}^{\rm true}$ as variable without further reference to
$\epsilon$. Doing so in the derivation of an expression for
the Ricci scalar $R[g]$, reshuffling partial $t$-derivatives,
and performing a similar ''regularization'' as
${\cal F}=0$,
we obtain precisely the expression for $R[g]$ as computed
within a manifestly ''continuous'' language, using the
''time''-coordinate $\tau$ (cf. the appendix of
Ref. \cite{KossowskiKriele1}), i.e. without $\delta$-function
terms. Proceeding with this definition
of the Ricci-scalar, and setting
$g^{1/2}=|\epsilon|^{1/2}\,N h^{1/2}$, one would end up again with
${\overline{\cal L}}$, hence the action $S_3$. This is another
example of the fact that the order in which one ''regularizes''
and sets $|\epsilon|=1$ may not be changed without changing the
resulting quantities as well. It illustrates that
not even the Ricci-scalar $R[g]$ is unambigously
well-defined if $K_{ij} \neq 0$.

In a manifestly ''continuous'' language,
these difficulties re-appear whenever
terms like $|\tau|^{1/2}$ have to be differentiated and
thereafter multiplied by singular objects (like $1/\tau$).
Formally using the product rule for derivatives,
one may generate additional $\delta$-functions easily.
Also, the rule $\partial_\tau(1/\tau)=-1/\tau^2$
is problematic in the sense of distributions if one
likes to look at $1/t^2$ as being the square of $1/t$.
Hence, reshuffling derivatives in the standard formulae
for the curvature quantities (which produces different
but equivalent versions of {\it one} equation in the usual case)
may give rise to truly different results when
$g_{\tau\tau}(\tau)$ has a zero. All these difficulties
show that the regular actions approach is better suited for
signature change.
\\
\\
\\
{\Large {\bf Acknowledgments}}
\medskip

I would like to acknowledge a few very good discussions with
Marcus Kriele on the {\it pros and cons}.
Moreover, he detected an error in the original manuscript.
Thanks also to Tevian Dray for pointing me out
that one may in principle admit a third version of signature
change (see Ref. \cite{Drayppp}).


\begin{thebibliography}{00}
\bibitem{Drayppp}
{
 I would like to thank T. Dray for pointing this out to me.
 The possibility of ''anticontinuous'' junction conditions
 is briefly mentioned in Ref. \cite{HellabyDray}, analogous
 junction conditions for a scalar field are considered in
 Refs. \cite{DrayManogueTucker1}, \cite{DrayManogueTucker2},
 and \cite{DrayManogueTucker3} (see also
 Ref. \cite{Hayward4}).
}
\end{thebibliography}
\end{document}